\documentclass{article}
\usepackage{amssymb}
\usepackage{amsfonts}
\usepackage{amsmath}
\usepackage{epsfig}

\input{tcilatex}
\begin{document}

\title{Neutrino tri-bi-maximal mixing from a non-Abelian discrete family
symmetry}
\author{I. de M. Varzielas\thanks{%
i.varzielas@physics.ox.ac.uk} $^{a}$, S. F. King\thanks{%
sfk@hep.phys.soton.ac.uk} $^{b}$, G. G. Ross\thanks{%
g.ross@physics.ox.ac.uk} $^{a}$ \\
$^{a}$ Rudolf Peierls Centre for Theoretical Physics,\\
University of Oxford, 1 Keble Road, Oxford, OX1 3NP\\
$^{b}$ CERN,1211 Geneva 23, Switzerland\\
and\\
School of Physics and Astronomy, \\
University of Southampton,\\
Southampton, SO17 1BJ, U.K.}
\maketitle

\begin{abstract}
The observed neutrino mixing, having a near maximal atmospheric neutrino
mixing angle and a large solar mixing angle, is close to tri-bi-maximal. We
argue that this structure suggests a family symmetric origin in which the
magnitude of the mixing angles are related to the existence of a discrete
non-Abelian family symmetry. We construct a model in which the family
symmetry is the non-Abelian discrete group $\Delta(27)$, a subgroup of $%
SU(3) $ in which the tri-bi-maximal mixing directly follows from the vacuum
structure enforced by the discrete symmetry. In addition to the lepton
mixing angles, the model accounts for the observed quark and lepton masses
and the CKM matrix. The structure is also consistent with an underlying
stage of Grand Unification.
\end{abstract}

\section{Introduction}

The observed neutrino oscillation parameters are consistent with a
tri-bi-maximal structure \cite{HPS}:%
\begin{equation}
U_{PMNS}\propto \left[ 
\begin{array}{ccc}
-\sqrt{\frac{2}{6}} & \sqrt{\frac{1}{3}} & 0 \\ 
\sqrt{\frac{1}{6}} & \sqrt{\frac{1}{3}} & \sqrt{\frac{1}{2}} \\ 
\sqrt{\frac{1}{6}} & \sqrt{\frac{1}{3}} & -\sqrt{\frac{1}{2}}%
\end{array}%
\right]  \label{eq:HPS}
\end{equation}

It has been observed that this simple form might be a hint of an underlying
family symmetry, and several models have been constructed that account for
this structure of leptonic mixing (e.g. \cite{Leptons}). It is possible to
extend the underlying family symmetry to provide a complete description of
the complete fermionic structure (e.g. \cite{Quarks}) \footnote{%
See \cite{Reviews} for review papers with extensive references on neutrino
models}, in which, in contrast to the neutrinos, the quarks have a strongly
hierarchical structure with small mixing with Yukawa coupling matrices of
the form \cite{Quark hierarchical ansatz}:

\begin{equation}
Y^{u}\propto \left[ 
\begin{array}{ccc}
0 & \epsilon _{u}^{3} & {O}\left( \epsilon _{u}^{3}\right) \\ 
\epsilon _{u}^{3} & \epsilon _{u}^{2} & {O}\left( \epsilon _{u}^{2}\right)
\\ 
{O}\left( \epsilon _{u}^{3}\right) & {O}\left( \epsilon _{u}^{2}\right) & 1%
\end{array}%
\right]  \label{eq:Yu}
\end{equation}

\begin{equation}
Y^{d}\propto \left[ 
\begin{array}{ccc}
0 & 1.5\epsilon _{d}^{3} & 0.4\epsilon _{d}^{3} \\ 
1.5\epsilon _{d}^{3} & \epsilon _{d}^{2} & 1.3\epsilon _{d}^{2} \\ 
O(\epsilon _{d}^{3}) & O(\epsilon _{d}^{2}) & 1%
\end{array}%
\right]  \label{eq:Yd}
\end{equation}%
where the expansion parameters are given by%
\begin{equation}
\epsilon_{u}\approx 0.05,\epsilon_{d}\approx 0.15.  \label{eq:u,d eps}
\end{equation}

A desirable feature of a complete model of quark and lepton masses and
mixing angles is that it should be consistent with an underlying Grand
Unified structure, either at the field theory level or at the level of the
superstring. The family symmetry models which have been built to achieve
this are based on an underlying $G_{f}\otimes SO(10)$ structure where the
family group $G_{f}$ is $SU(3)_{f}\cite{Ivo SU(3),King and Ross SU(3) model}%
. $ This is very constraining because it requires that all the (left handed)
members of a single family should have the same family charge. In this paper
we will construct a model based on a non-Abelian discrete family symmetry
which preserves the possibility of simple unification by requiring that the
discrete symmetry properties of all the members of one family are the same.
The discrete non-Abelian group\footnote{%
Such non-Abelian discrete symmetries often occur in compactified string
models.} we use is $\Delta (27),$ the semi-direct product group $%
Z_{3}\ltimes Z_{3}^{\prime }$, which is a subgroup \footnote{$Z_{3}\ltimes
Z_{3}^{\prime }$ (where the generators of the distinct $Z_{3}$ don't
commute) is the group $\Delta (27)$ \cite{FFK}} of $SU(3)_{f}.$ Indeed the
dominant terms of the Lagrangian leading to the Yukawa coupling matrices of
the form of eq.(\ref{eq:Yu}) and eq.(\ref{eq:Yd}) are symmetric under $%
SU(3)_{f}$ so much of the structure of the model based on $SU(3)_{f}$ is
maintained. However the appearance of additional terms allowed by $%
Z_{3}\ltimes Z_{3}^{\prime }$ but not by $SU(3)_{f}$ determines the vacuum
structure and generates the tri-bi-maximal mixing structure. The choice of
the multiplet structure ensures that the model is consistent with a stage of
Grand or superstring unification and the resulting model is much simpler
than that based on the continuous $SU(3)_{f}$ $\ $symmetry.

In Section \ref{sec:Fields and symmetries} we discuss the choice of the
non-Abelian discrete group and the multiplet content of the model. Emphasis
is put on obtaining a simplified field content and a reduced auxiliary
symmetry compared with the $SU(3)_{f}$ model in \cite{Ivo SU(3)}. In Section %
\ref{sec:VEVs and masses} we consider the superpotential terms allowed by
the symmetries of the model. Using this we show how the desired vacuum
structure arises simply through the appearance of the additional invariants
allowed by $Z_{3}\ltimes Z_{3}^{\prime }$ but not by $SU(3)_{f}$. Section %
\ref{sub:Yukawa} discusses both the Dirac and Majorana mass matrix structure
of the model and the resulting pattern of quark, charged lepton and neutrino
masses and mixing angles. Finally in Section \ref{sec:Conclusion} we present
a summary and our conclusions.

\section{Field content and symmetries\label{sec:Fields and symmetries}}

The symmetry of the model is $G_{f}\otimes SU(3)\otimes SU(2)\otimes
U(1)\otimes G_{f}\otimes G.$ The additional symmetry group $G$ is needed to
restrict the form of the allowed coupling of the theory and is chosen to be
as simple as possible. As discussed above, the family group $G_{f}$ is
chosen as a non-Abelian discrete group of $SU(3)_{f}$ in a manner that
preserves the structure of the fermion Yukawa couplings of the associated $%
SU(3)_{f}$ model of \cite{Ivo SU(3)}. This means that $G_{f}$ should be a
non-Abelian subgroup of $SU(3)_{f}$ of sufficient size that it approximates $%
SU(3)_{f}$ in the sense that most of the leading terms responsible for the
fermion mass structure in the $SU(3)_{f}$ are still the leading terms
allowed by $G_{f}$ (which being a subgroup, allows further terms which we
want to be subleading). The smallest group we have found that achieves this
is $\Delta (27),$ the semi-direct product group $Z_{3}\ltimes Z_{3}^{\prime
} $. The main change that results from using this smaller symmetry group is
the appearance of additional invariants which drive the desired vacuum
structure and, because we are no longer dealing with a continuous symmetry,
the absence of the associated $D$-terms which were very important in
determining the vacuum structure in the $SU(3)_{f}$ model \cite{Ivo SU(3)}.
Due to this, we are able to reduce the total field content of this model,
which in turn only requires an additional $G=U(1)\otimes Z_{2}\otimes R$ to
control the allowed terms in the superpotential \footnote{$R$ is an $R-$
symmetry and for SUSY purposes plays the same role as $R-$parity.} (\textit{%
c.f.} $U(1)\otimes U(1)^{\prime }$ $\otimes R$ in \cite{Ivo SU(3)}).

In choosing the representation content of the theory we are guided by the
structure of the $SU(3)_{f}$ model of \cite{Ivo SU(3)} which generated a
viable form of all quark and lepton masses and mixing. Since $Z_{3}\ltimes
Z_{3}^{\prime }$ is a discrete subgroup of $SU(3)_{f}$ all invariants of $%
SU(3)_{f}$ are invariants of $Z_{3}\ltimes Z_{3}^{\prime }.$ Using this we
can readily arrange that the superpotential terms responsible for fermion
masses in the $SU(3)_{f}$ model are present in the $Z_{3}\ltimes
Z_{3}^{\prime }$ model. To implement this we find it convenient to label the
representation of the fields of our model by their transformation properties
under the approximate $SU(3)_{f}$ family group. The Standard Model (SM)
fermions $\psi _{i},\psi _{j}^{c}$ transform as triplets under this group.
The transformation properties of such triplets under the $Z_{3}\ltimes
Z_{3}^{\prime }$ discrete group are shown in Table \ref{transformation}.
Although the gauge group is just that of the Standard Model it is also
instructive, in considering how the model can be embedded in a unified
structure, to display the properties of the states under the $%
SU(4)_{PS}\otimes SU(2)_{L}\otimes SU(2)_{R}$ subgroup of $SO(10)$ and this
is done in Table \ref{fields}$.$ We also show in Table \ref{fields} the
transformation properties under the additional symmetry group $G= U(1)
\otimes Z_{2} \otimes R$. The transformation properties of the SM Higgs, $H$%
, responsible for electroweak breaking \footnote{%
Two Higgs are required due to SUSY, represented as $H$, they have the same
charges under $G_{f}$ and $G$.} are also shown in Table \ref{fields}.

In a complete unified theory, quark and lepton masses will be related. A
particular question that arises in such unification is what generates the
difference between the down quark and charged lepton masses. In \cite{Ivo
SU(3)} this was done through a variant of the Georgi-Jarlskog mechanism \cite%
{GeorgiJarlskog} via the introduction of another Higgs field $H_{45}$, which
transforms as a $45$ of an underlying $SO(10)$ GUT. It has a vacuum
expectation value (vev) which breaks $SO(10)$ but leaves the SM gauge group
unbroken. In this model we include $H_{45}$ to demonstrate that the model
readily Grand Unifies but in practice we only use its vev. This does not
necessarily imply that there is an underlying stage of Grand Unification
below the string scale but, if not, the underlying theory should provide an
alternative explanation for the existence of the pattern of low energy
couplings implied by terms involving $H_{45}.$

At this stage there are no terms generating fermion masses and to complete
the model it is necessary to break the family symmetry $Z_{3}\ltimes
Z_{3}^{\prime }$ through the introduction of {}\textquotedblleft
flavons\textquotedblright\ that acquire vevs. To reproduce the results of
the phenomenologically viable $SU(3)_{f}$ model \cite{Ivo SU(3)} we choose a
similar but somewhat simplified flavon structure with the $SU(3)_{f}$
antitriplet fields $\theta ^{i},$ $\bar{\phi}_{3}^{i}$, $\bar{\phi}_{23}^{i}$
and $\bar{\phi}_{123}^{i}$ and $SU(3)_{f}$ triplet fields $\phi _{3_{i}}$, $%
\phi _{3_{i}}^{\prime }$ as shown in Table \ref{fields}, and one triplet
field for alignment purposes $\phi _{i}^{A}$. The transformation properties
of these fields under $Z_{3}\ltimes Z_{3}^{\prime }$ are shown in Table \ref%
{transformation}. With this choice, as discussed in the next Section, the
Yukawa structure of the $SU(3)_{f}$ model \cite{Ivo SU(3)} is obtained. One
may readily check that the additional terms allowed by the $Z_{3}\ltimes
Z_{3}^{\prime }$ symmetry are subleading in this sector so the
phenomenologically acceptable pattern of fermion masses and mixings obtained
in \cite{Ivo SU(3)} is reproduced here if the flavon vacuum structure is as
given in \cite{Ivo SU(3)}. The main difference between the models is the
appearance in the potential determining the vacuum structure of additional
invariants allowed by $Z_{3}\ltimes Z_{3}^{\prime }$ and the absence of the $%
D-$terms associated with a continuous gauge symmetry.

\begin{table}[tbp] \centering%
\begin{tabular}{|c||c|c|}
\hline
Field & $Z_{3}$ & $Z_{3}^{\prime }$ \\ \hline
$\phi_{1}$ & $\phi_{1}$ & $\phi_{2}$ \\ 
$\phi_{2}$ & $\alpha \phi_{2}$ & $\phi_{3}$ \\ 
$\phi_{3}$ & $\left( \alpha \right) ^{2}\phi_{3}$ & $\phi_{1}$ \\ 
$\bar{\phi} ^{1}$ & $\bar{\phi} ^{1}$ & $\bar{\phi} ^{2}$ \\ 
$\bar{\phi} ^{2}$ & $\left( \alpha \right) ^{2}\bar{\phi} ^{2}$ & $\bar{\phi}%
^{3}$ \\ 
$\bar{\phi} ^{3}$ & $\alpha \bar{\phi} ^{3}$ & $\bar{\phi} ^{1}$ \\ \hline
\end{tabular}%
\caption{Transformation properties of $SU(3)_{f}$ anti-triplet fields $\bar{\phi}^{i}$
and triplet fields $\phi_i$ under the non-Abelian discrete group; $\alpha$ is the cube root of unity,
$\alpha^{3}=1$.}\label{transformation}%
\end{table}%

\begin{table}[tbp] \centering%
\begin{tabular}{|c||c||c|c|c||c||c|c|}
\hline
Field & $SU(3)_{f}$ & $SU(4)_{PS}$ & $SU(2)_{L}$ & $SU(2)_{R}$ & $R$ & $U(1)$
& $Z_{2}$ \\ \hline\hline
$\psi $ & $\mathbf{3}$ & $\mathbf{4}$ & $\mathbf{2}$ & $\mathbf{1}$ & $%
\mathbf{1}$ & $\mathbf{0}$ & $\mathbf{1}$ \\ 
$\psi ^{c}$ & $\mathbf{3}$ & $\bar{\mathbf{4}}$ & $\mathbf{1}$ & $\mathbf{2}$
& $\mathbf{1}$ & $\mathbf{0}$ & $\mathbf{1}$ \\ 
$\theta $ & $\bar{\mathbf{3}}$ & $\mathbf{4}$ & $\mathbf{1}$ & $\mathbf{2}$
& $\mathbf{0}$ & $\mathbf{0}$ & $-\mathbf{1}$ \\ \hline
$H$ & $\mathbf{1}$ & $\mathbf{1}$ & $\mathbf{2}$ & $\mathbf{2}$ & $\mathbf{0}
$ & $\mathbf{0}$ & $\mathbf{1}$ \\ 
$H_{45}$ & $\mathbf{1}$ & $\mathbf{15}$ & $\mathbf{1}$ & $\mathbf{3}$ & $%
\mathbf{0}$ & $\mathbf{2}$ & $\mathbf{1}$ \\ \hline\hline
$\phi _{123}$ & $\mathbf{3}$ & $\mathbf{1}$ & $\mathbf{1}$ & $\mathbf{1}$ & $%
\mathbf{0}$ & $\mathbf{-1}$ & $\mathbf{1}$ \\ 
$\phi _{3}$ & $\mathbf{3}$ & $\mathbf{1}$ & $\mathbf{1}$ & $\mathbf{1}$ & $%
\mathbf{0}$ & $\mathbf{3}$ & $\mathbf{1}$ \\ 
$\phi_{1}$ & $\mathbf{3}$ & $\mathbf{1}$ & $\mathbf{1}$ & $\mathbf{1}$ & $%
\mathbf{0}$ & $-\mathbf{4}$ & $-\mathbf{1}$ \\ 
$\bar{\phi}_{3}$ & $\mathbf{\bar{\mathbf{3}}}$ & $\mathbf{1}$ & $\mathbf{1}$
& $\mathbf{3}\oplus \mathbf{1}$ & $\mathbf{0}$ & $\mathbf{0}$ & $-\mathbf{1}$
\\ 
$\bar{\phi}_{23}$ & $\mathbf{\bar{\mathbf{3}}}$ & $\mathbf{1}$ & $\mathbf{1}$
& $\mathbf{1}$ & $\mathbf{0}$ & $-\mathbf{1}$ & $-\mathbf{1}$ \\ 
$\bar{\phi}_{123}$ & $\mathbf{\bar{\mathbf{3}}}$ & $\mathbf{1}$ & $\mathbf{1}
$ & $\mathbf{1}$ & $\mathbf{0}$ & $\mathbf{1}$ & $-\mathbf{1}$ \\ \hline
\end{tabular}%
\caption{Symmetries and Charges}\label{fields}%
\end{table}%

\section{Symmetry breaking \label{sec:VEVs and masses}}

Following \cite{Ivo SU(3)} the desired pattern of vevs is given by

\begin{equation}
\left\langle \bar{\phi}_{3}\right\rangle ^{T}=\left(%
\begin{array}{c}
0 \\ 
0 \\ 
1%
\end{array}%
\right)\otimes\left(%
\begin{array}{cc}
a_{u} & 0 \\ 
0 & a_{d}%
\end{array}%
\right)  \label{eq:P3 vev}
\end{equation}

\begin{equation}
\left\langle \bar{\phi}_{23}\right\rangle ^{T}=\left(%
\begin{array}{c}
0 \\ 
-b \\ 
b%
\end{array}%
\right)  \label{eq:P23 vev}
\end{equation}

\begin{equation}
\left\langle \phi_{123} \right\rangle \propto \left\langle \bar{\phi}%
_{123}\right\rangle ^{T}=\left( 
\begin{array}{c}
c \\ 
c \\ 
c%
\end{array}%
\right)  \label{eq:P123 vev}
\end{equation}

\begin{equation}
\left\langle \phi_{1} \right\rangle \propto \left( 
\begin{array}{c}
1 \\ 
0 \\ 
0%
\end{array}%
\right)  \label{eq:P1 vev}
\end{equation}

\begin{equation}
\left\langle \theta \right\rangle \propto \left\langle \phi_{3}\right\rangle
\propto \left( 
\begin{array}{c}
0 \\ 
0 \\ 
1%
\end{array}%
\right)  \label{eq:T vev}
\end{equation}%
where the $SU(2)_{R}$ structure of $\left\langle \bar{\phi}
_{3}\right\rangle $ has been displayed.

The alignment of these vevs can proceed in various ways. By including
additional driving fields in the manner discussed in \cite{GIS on subgroups}
one can arrange their $F-$terms give a scalar potential whose minimum has
the desired vacuum alignment. Here however we show that an even simpler
mechanism involving $D-$terms only achieves the desired alignment.

To understand how this vacuum alignment works note that, unlike the case for
the continuous $SU(3)_{f}$ symmetric theory, it is not possible in general
to rotate the vacuum expectation value of a triplet field to a single
direction, for example the $3$ direction. Due to the underlying discrete
symmetry the vev will be quantised in one of a finite set of possible
minima. However this may only be apparent if higher order terms in the
potential are included for the lower order terms may have the enhanced $%
SU(3)_{f}$ symmetry.

To make this more explicit, consider a general $SU(3)_{f}$ triplet field $%
\phi _{i}$. It will have a SUSY breaking soft mass term in the Lagrangian of
the form $m_{\phi }^{2}\phi ^{i^{\dagger }}\phi _{i}$ which is invariant
under the approximate $SU(3)_{f}$ symmetry. Radiative corrections involving
superpotential couplings to massive states may drive the mass squared
negative at some scale $\Lambda $ triggering a vev for the field $\phi ,$ $%
<\phi ^{i^{\dagger }}\phi _{i}>=v^{2},$ with $v^{2}\leq \Lambda ^{2}$ set
radiatively \footnote{%
The radiative corrections to the soft mass term depend on the details of the
underlying theory at the string or unification scale.}. At this stage the
vev of $\phi $ can always be rotated to the $3$ direction using the
approximate $SU(3)_{f}$ symmetry. However this does not remain true when
higher order terms allowed by the discrete family symmetry are included. For
the model considered here the leading higher order term is of the form $%
m_{3/2}^{2}(\phi ^{\dagger }\phi \phi ^{\dagger }\phi )$ arising as a
component of the $D-$ term $\left[ \chi ^{\dag }\chi (\phi ^{\dagger }\phi
\phi ^{\dagger }\phi )\right] _{D}$. In this we have suppressed the coupling
constants and the messenger mass scale (or scales), $M,$ associated with
these higher dimension operators (which can even be the Planck mass $M_{P}$%
). The $F$ component of the field $\chi $ drives supersymmetry breaking and $%
m_{3/2}$ is the graviton mass ($m_{3/2}^{2}=F_{\chi }^{\dagger }F_{\chi
}/M_{P}^{2}).$ This term gives rise to two independent quartic invariants
under $Z_{3}\ltimes Z_{3}^{\prime },$ namely $m_{3/2}^{2}(\phi ^{i^{\dagger
}}\phi _{i}\phi ^{j^{\dagger }}\phi _{j})$ and $m_{3/2}^{2}(\phi
^{i^{\dagger }}\phi _{i}\phi ^{i^{\dagger }}\phi _{i}).$ The former is $%
SU(3)_{f}$ symmetric and does not remove the vacuum degeneracy. The second
term is not $SU(3)_{f}$ symmetric and does lead to an unique vacuum state.
For the case that the coefficient of $m_{3/2}^{2}(\phi ^{i^{\dagger }}\phi
_{i}\phi ^{i^{\dagger }}\phi _{i})$ is positive the minimum corresponds to
the vev \footnote{%
In general, the phases are different for each entry of this vev. For
simplicity we omit them, as they don't affect the results.} $<\phi
_{i}>^{T}=v(1,1,1)/\sqrt{3}$ (c.f. eq.(\ref{eq:P123 vev})). For the case the
coefficient is negative, the vev has the form $<\phi _{i}>^{T}=v(0,0,1)$
(c.f. eq.(\ref{eq:T vev})). Thus we see that, in contrast to the continuous
symmetry case, the discrete non-Abelian symmetry leads to a finite number of
candidate vacuum states. Which one is chosen depends on the sign of the
higher dimension term which in turn depends on the details of the underlying
theory. In this paper we do not attempt to construct the full theory and so
cannot determine this sign. What we will demonstrate, however, is that one
of the finite number of candidate vacua does have the correct properties to
generate a viable theory of fermion masses and mixings.

The vacuum alignment needed for this model can now readily be obtained.
Suppose that a combination of radiative corrections and the $U(1)$ $D$-term
drive $m_{\phi _{123}}^{2}$, $m_{\phi _{1}}^{2}$ and $m_{\bar{\phi}_{3}}^{2}$
negative close to the messenger scale, $\Lambda _{\phi _{123},\phi
_{1}},_{\bar{\phi} _{3}}\lesssim M$. The symmetries of the model ensure that the
leading terms fixing their vacuum structure are of the form $%
m_{3/2}^{2}(\phi _{123}^{\dagger }\phi _{123}\phi _{123}^{\dagger }\phi
_{123})$, $m_{3/2}^{2}(\phi _{1}^{\dagger }\phi _{1}\phi _{1}^{\dagger }\phi
_{1}),$ $m_{3/2}^{2}(\phi _{123}^{\dagger }\phi _{123}\phi _{1}^{\dagger
}\phi _{1}),$ plus similar terms involving $\bar{\phi} _{3}$. Provided the unmixed
terms of the form of the first two terms dominate the vevs will be
determined by the signs of these terms. If the quartic term involving $\phi
_{123}$ is positive $\phi _{123}$ will acquire a vev in the $(1,1,1)$
direction as in eq.(\ref{eq:P123 vev}). If the quartic term involving $\phi
_{1}$ is negative $\phi _{1}$ will acquire a vev in the $(1,0,0)$ direction
as in eq.(\ref{eq:P1 vev}) where the non zero entry just defines the $1$
direction. Finally if the quartic term involving $\bar{\phi} _{3}$ is also
negative it will acquire a vev with a single non-zero entry but the position
of this entry will depend on the leading $D-$term resolving this ambiguity.
If the term $m_{3/2}^{2}(\bar{\phi}^{i} _{3} \phi _{1_{i}} \phi _{1}^{\dagger j
}\bar{\phi} _{3_{j}}^{\dagger })$ dominates and has positive coefficient it will force the vevs
of these fields to be orthogonal and so $\bar{\phi} _{3}$ has a vev in the $(0,0,1)
$ direction, c.f. eq.(\ref{eq:P3 vev}), where again the non zero entry just
defines the $3$ direction. In a similar manner it is straightforward to see
how the fields $\phi_{3}$ and $\theta $ align along the $(0,0,1)$
direction if the quartic terms $m_{3/2}^{2}(\bar{\phi}^{i}_{3} \phi_{3_{i}} \phi _{3}^{\dagger j}\bar{\phi} _{3_{j}}^{\dagger })$ and
$m_{3/2}^{2}(\bar{\phi}^{i}_{3} \theta_{i} \theta ^{\dagger j}\bar{\phi }^{\dagger }_{3_{j}} )$ dominate and have
negative coefficients. The scale of their vevs is determined by the scale at
which their soft mass squared become negative (the direction of $\langle \phi _{3}\rangle $ is
not very relevant, but with the above terms similar to $\theta$ it can take the form in eq.(%
\ref{eq:T vev}) and we take it to be so for simplicity).

The relative alignment of the remaining terms follows in a similar manner.
Consider the field $\bar{\phi}_{23}$ with a soft mass squared becoming
negative at a scale $b<v$. For $\bar{\phi}_{23}$
we want the dominant term aligning its vev to be $m_{3/2}^{2}(\bar{\phi}%
_{23}^{i}\phi _{123_{i}}\phi _{123}^{\dagger j}\bar{\phi}_{23_{j}}^{\dagger
})$, with positive coefficient. It will then acquire a vev orthogonal to
that of $\phi _{123}$. The choice of the particular orthogonal direction
will be determined by terms like $m_{3/2}^{2}(\bar{\phi}_{3}^{i}\bar{\phi}%
_{23_{i}}^{\dagger }\bar{\phi}_{23}^{i}\bar{\phi}_{3_{i}}^{\dagger })$ or $%
m_{3/2}^{2}(\bar{\phi}_{23}^{i}\phi _{1_{i}}\phi _{1}^{\dagger j}\bar{\phi}%
_{23_{j}}^{\dagger })$ . If the latter dominates with a positive
coefficient, it will drive $\langle \bar{\phi}_{23}\rangle $ orthogonal to $%
\phi _{1}$ - the form given in eq.(\ref{eq:P23 vev}).

Finally consider the field $\bar{\phi}_{123}$ with a soft mass squared
becoming negative at a scale $c\ll v.$ The leading terms determining its
vacuum alignment are $m_{3/2}^{2}(\bar{\phi}_{3}^{i}\bar{\phi}%
_{23_{i}}^{\dagger }\bar{\phi}_{3}^{j}\bar{\phi}_{123_{j}}^{\dagger })$ and $%
m_{3/2}^{2}(\bar{\phi}_{123}^{i}\phi _{123_{i}}\phi _{123}^{\dagger j}\bar{%
\phi}_{123_{j}}^{\dagger })$ . If the latter dominates with a negative
coefficient, $\bar{\phi}_{123}$ will be aligned in the same direction as $%
\phi _{123}$ and have the form given in eq.(\ref{eq:P123 vev}). Note that
the term involving $\bar{\phi}_{23}$ is accidental in the sense that it is
dependant on the $U(1)$ assignments of the field.

In summary, we have shown that higher order $D-$terms constrained by the
discrete family symmetry lead to a discrete number of possible vacuum
states. Which one is the vacuum state depends on the coefficients of these
higher order terms which are determined by the underlying unified GUT or
string theory. Our analysis has shown that the vacuum structure needed for a
viable theory of fermion masses can readily emerge from this discrete set of
states.

\section{The mass matrix structure\label{sub:Yukawa}}

\subsection{Yukawa terms}

We turn now to the structure of the quark and lepton mass matrices. The
leading Yukawa terms allowed by the symmetries are:

\begin{equation}
P_{Y}\sim\frac{1}{M^{2}}\bar{\phi}_{3}^{i}\psi_{i}\bar{\phi}%
_{3}^{j}\psi_{j}^{c}H  \label{eq:Y_P3_P3}
\end{equation}
\begin{equation}
+\frac{1}{M^{3}}\bar{\phi}_{23}^{i}\psi_{i}\bar{\phi}_{23}^{j}%
\psi_{j}^{c}HH_{45}  \label{eq:Y_P23_P23}
\end{equation}

\begin{equation}
+\frac{1}{M^{2}}\bar{\phi}_{23}^{i}\psi_{i}\bar{\phi}_{123}^{j}\psi_{j}^{c}H
\label{eq:Y_P23_P123}
\end{equation}
\begin{equation}
+\frac{1}{M^{2}}\bar{\phi}_{123}^{i}\psi_{i}\bar{\phi}_{23}^{j}\psi_{j}^{c}H
\label{eq:Y_P123_P23}
\end{equation}

\begin{equation}
+\frac{1}{M^{5}}\bar{\phi}_{123}^{i}\psi_{i}^{c}\bar{\phi}%
_{3}^{j}\psi_{j}^{c}H H_{45}\bar{\phi}_{123}^{k}\phi_{1_{k}}
\label{eq:Y_P123_P3}
\end{equation}

\begin{equation}
+\frac{1}{M^{5}}\bar{\phi}_{3}^{i}\psi_{i}^{c}\bar{\phi}_{123}^{j}%
\psi_{j}^{c}H H_{45}\bar{\phi}_{123}^{k}\phi_{1_{k}}  \label{eq:Y_P3_P123}
\end{equation}

\begin{equation}
+\frac{1}{M^{6}}\bar{\phi}_{123}^{i}\psi_{i}^{c}\bar{\phi}%
_{123}^{j}\psi_{j}^{c}H\bar{\phi}_{3}^{k}\phi_{123_{k}}\bar{\phi}%
_{3}^{l}\phi_{123_{l}}  \label{eq:Y_P123_P123}
\end{equation}

Although of a slightly different from from that in \cite{Ivo SU(3)} these
terms realize the same mass structure and we refer the reader to \cite{Ivo
SU(3)} for the details. It gives a phenomenologically consistent description
of all the quark masses and mixing angles and the charged lepton masses,
generating their hierarchical structure through an expansion in the family
symmetry breaking parameters. The main differences in the way this is
achieved lies in eqs. (\ref{eq:Y_P123_P3}, \ref{eq:Y_P3_P123}, \ref%
{eq:Y_P123_P123}). The terms in eqs. (\ref{eq:Y_P123_P3},\ref{eq:Y_P3_P123})
account for the observed $O\left( \epsilon _{d}^{3}\right) $ difference in
the $12,21$ and $13,31$ entries \footnote{%
We take a symmetric form for the mass matrices as would be expected if there
is an underlying $SO(10)$ GUT \cite{Ivo SU(3)}} of the down-type quark mass
matrix (c.f. eq.(\ref{eq:Yd})) \cite{Quark hierarchical ansatz}.

The term in eq.(\ref{eq:Y_P123_P123}) is undesirable, but allowed by the
symmetries nonetheless. Naively, one expects it would contribute to the $11$
element at $O\left( \epsilon _{d}^{4}\right) $ giving unwanted corrections
to the phenomenologically successful Gatto-Sartori-Tonin relation \cite{GST}
which results if the $11$ entry is less than this order \cite{Ivo SU(3)}.
Fortunately, this texture zero is preserved at that order, as the vevs of $%
\phi _{3}$ and $\bar{\phi}_{3}$ are slightly smaller than the relevant
messenger mass scales, and in the eq.(\ref{eq:Y_P123_P123}) there are four
such fields, suppressing the term sufficiently. As such, the desired small
magnitude of this term can be maintained while keeping the dimensionless
coefficients in front of all the allowed Yukawa terms as $O(1)$.

\subsection{Majorana terms\label{sub:Majorana}}

The leading terms that contribute to the right-handed neutrino Majorana
masses are:

\begin{equation}
P_{M}\sim \frac{1}{M}\theta ^{i}\psi_{i}^{c}\theta ^{j}\psi_{j}^{c}
\label{eq:M_th_th}
\end{equation}%
\begin{equation}
+\frac{1}{M^{5}}\bar{\phi}_{23}^{i}\psi_{i}^{c}\bar{\phi}_{23}^{j}\psi
_{j}^{c}\theta ^{k}\phi_{123_{k}}\theta ^{l}\phi_{3_{l}}
\label{eq:M_P23_P23}
\end{equation}%
\begin{equation}
+\frac{1}{M^{5}}\bar{\phi}_{123}^{i}\psi_{i}^{c}\bar{\phi}_{123}^{j}\psi
_{j}^{c}\theta ^{k}\phi_{123_{k}}\theta ^{l}\phi_{123_{l}}
\label{eq:M_P123_P123}
\end{equation}

Note that these terms are different from those in \cite{Ivo SU(3)} and lead
to a different form for the ratios of the Majorana masses. The vev of $\phi
_{3}$ controls the hierarchy between $M_{1}$ (given essentially by eq.(\ref%
{eq:M_P123_P123})) and $M_{2}$ (from eq.(\ref{eq:M_P23_P23})). It is set by
radiative breaking to lie close to the scale of $<\bar{\phi} _{23}>$, such that
after seesaw we can fit the ratio of the neutrino squared mass differences $%
\frac{\Delta m_{\odot }^{2}}{\Delta m_{@}^{2}}$ \footnote{%
This is different from the $SU(3)_{f}$ model \cite{Ivo SU(3)} which
predicted the ratio $\frac{M_{1}}{M_{2}}$ to be associated with the
expansion parameter $\epsilon _{d}$ that was set by the quark sector, and
was consistent with the experimentally measured value $\frac{\Delta m_{\odot
}^{2}}{\Delta m_{@}^{2}}$.}. The hierarchy between the lightest Majorana
mass $M_{1}$, and the heaviest, $M_{3}$ is 
\begin{equation}
\frac{M_{1}}{M_{3}}\simeq \epsilon _{d}^{4}\frac{M_{d}^{4}}{M_{\nu _{R}}^{4}}
\label{eq:M1 M3}
\end{equation}%
where $M_{d}$ is the mass of the messenger responsable for the down quark
mass (for details on the messenger sector, we again refer the reader to \cite%
{Ivo SU(3)}).

For a viable pattern of neutrino mixing we need to ensure that the hierarchy
in eq.(\ref{eq:M1 M3}) is sufficiently strong to suppress the contribution
from $\nu _{3}^{c}$ exchange which would otherwise give an unacceptably
large $\nu _{\tau }$ component in the atmospheric (and/or solar) neutrino
eigenstates. This requirement on the Majorana hierarchy puts a lower bound
on the mass of corresponding right-handed neutrino messenger, as is clear
from eq.(\ref{eq:M1 M3}). The light neutrino eigenstates also have an
hierarchical mass structure so the heaviest of the light effective neutrinos
has a mass given approximately by $\sqrt{\Delta m_{@}^{2}}$. Using this,
together with eq.(\ref{eq:M1 M3}), we find

\begin{equation}
M_{3} \simeq \epsilon _{d}^{2}\langle H \rangle ^{2}\frac{M_{\nu _{R}}^{4}}{%
M_{\nu }^{4}}\Delta m_{@}^{2}{}^{-\frac{1}{2}}\simeq 10^{13}\frac{M_{\nu
_{R}}^{4}}{M_{\nu }^{4}}GeV  \label{eq:M3 bound}
\end{equation}%
where $M_{\nu }$ is the mass of the messenger responsable for the Dirac
neutrino mass. \ 

The final structure of neutrino mixing is very similar to the one in \cite%
{Ivo SU(3)}, and generates the same predictions for the neutrino mixing
angles. The leptonic mixing angles are obtained after taking into account
the (small) effect of the charged leptons, yielding nearly tri-bi-maximal
mixing \cite{Antusch}:

\begin{equation}
\sin^{2}\theta_{12} = \frac{1}{3}\pm_{0.048}^{0.052}
\end{equation}

\begin{equation}
\sin^{2}\theta_{23} = \frac{1}{2}\pm_{0.058}^{0.061}
\end{equation}

\begin{equation}
\sin^{2}\theta_{13} = 0.0028
\end{equation}

This leads to the prediction for the reactor angle of $\theta_{13} \approx
\theta_C /(3\sqrt{2}) \approx 3^o$, where $\theta_C$ is the Cabibbo angle,
i.e. the prediction is a factor of 3 smaller than the Cabibbo angle due to
the Georgi-Jarlskog factor, and also a factor of $\sqrt{2}$ smaller due to
commutation through the maximal atmospheric angle. Also $\theta_{12}$ can be
related to $\theta_{13}$ and the CP violating phase $\delta$, via the so
called neutrino sum rule first derived by one of us in \cite{Quarks}: 
\begin{equation}
\theta_{12}+\theta_{13}\cos(\delta -\pi)\approx 35.26^o.
\end{equation}
The above predictions were first shown to follow from the charged lepton
corrections to tri-bi-maximal mixing in the $SO(3)$ model proposed by one of
us in \cite{Quarks} and later shown to be applicable to a class of models in 
\cite{Antusch}, including the present model discussed here and in \cite{Ivo
SU(3)} \footnote{%
Note that the prediction for $\theta_{13}$ in \cite{Ivo SU(3)} has been
corrected here.}.

\section{Summary and conclusions\label{sec:Conclusion}}

We have constructed a complete theory of fermion masses and mixings based on
the spontaneous breaking of the discrete non-Abelian symmetry group $%
Z_{3}\ltimes Z_{3}^{\prime }.$ The model is constructed in a manner
consistent with an underlying Grand Unified symmetry with all the members of
a family of fermions having the same symmetry properties under the family
symmetry group. Many of the properties of the model rely on the approximate $%
SU(3)_{f}$ symmetry that the discrete group possesses and the model is very
close to the continuous $SU(3)_{f}$ family symmetry model of reference \cite%
{Ivo SU(3)}. The main difference is a significant simplification in the
vacuum alignment mechanism in which the near tri-bi-maximal mixing in the
lepton sector directly follows from the non-Abelian discrete group. In
addition to the prediction of near tri-bi-maximal mixing the model preserves
the Gatto-Sartori-Tonin \cite{GST} relation between the light quark masses
and the Cabibbo mixing angle, and can accommodate the GUT relations between
the down quark and lepton masses. It also provides a explanation for the
hierarchy of quark masses and mixing angles in terms of an expansion in
powers of a family symmetry breaking parameter.

\section*{Acknowledgments}
We are grateful to Michal Malinsky for pointing out an error
in the vacuum alignment discussion in the original version
of this paper. 

\noindent The work of I. de M. Varzielas was supported by FCT under the grant
SFRH/BD/12218/2003.

\noindent This work was partially supported by the EC 6th Framework
Programme MRTN-CT-2004-503369.

\end{document}